# The three-dimensional structure of population density in world cities

Gaëtan LAZIOU[a,*] and Rémi LEMOY[a]

[a]University of Rouen, IDEES Laboratory UMR 6266 CNRS, Mont-Saint-Aignan, France

---

* Corresponding author:

Gaëtan Laziou

E-mail address: gaetan.laziou1@univ-rouen.fr

# Abstract

A good understanding of cities is crucial to implement urban planning policies leading to social and economic sustainability and an efficient use of resources. While urban concentration has been associated with both positive and negative effects, echoing debates on compact cities, few studies have documented how density evolves with city size. We fill this gap by investigating how the population density radial structure changes across the urban hierarchy. Our results uncover strong regularities in urban settlements. In terms of density, cities can be seen as exponential cones which evolve homothetically with city population. This rather simple but universal geometric structure of cities provides a new spatial scaling law, which is an important step forward in understanding how cities work and grow. Some deviations can be observed, which mainly oppose dense cities in the developing world and sprawled cities in high-income countries, associated with high energy use per capita. This suggests that urban lifestyle in wealthiest countries has come at the price of negative impacts on environmental outcomes. This research has a broad range of applications as it provides a powerful tool to compare cities of different sizes.

# Introduction

Although they cover a small share of the land area, cities have high population densities and concentrate more than one half of the world's population[1,2]. Therefore, urban density is key in understanding modern lifestyles. Further, it has many implications on various facets of life, as housing, transport or air quality. Studies by physicists, geographers and economists have associated density with positive effects, for example increased productivity and more efficient public transports, but also negative effects, as high housing costs and traffic congestion[3–7]. While this strand of the literature is widely developed, the precise description of the urban structure and how it evolves as we move from a small to a large city, has received less attention. This is a shortcoming, because it is crucial to understand if similar mechanisms drive the internal structure of cities, in order to make them more sustainable and implement appropriate planning policies.

A milestone in the study of population density structure is the seminal work of Clark (1951)[8], later grounded theoretically[9]. His empirical research confirms earlier observation[10], by showing that population density tends to decrease exponentially with the distance from the center in many cities around the world. Since then, many authors have used a negative exponential function to model population density within cities[11–13], but other functional forms have also been studied[14–16]. Some studies have proposed comprehensive reviews of these population density functions[17,18]. In recent years and mostly driven by the growing availability of data, these empirical analyses have benefited from a renewed interest. Most of this literature focuses on change over time in the distribution of population[19–21].

This modeling effort has also been followed by several attempts to compare cities with each other, and explain variations between them. From the beginning, it has been suggested that transport costs, income, city size and the development of public transportation have an impact on urban density, although no formal analysis supported this claim[8]. Later work found that the density profile is flatter when income is higher and transport costs lower[13]. Besides, there is a consensus on the fact that the urban density radial structure is influenced by city size. Numerous studies in the literature highlight that both density decay and central density are dependent on city population[11,22–24]. Building on this observation, some authors have attempted to remove the city size effect in order to compare population concentration between cities[21].

The scaling framework, on the other hand, has developed independently from this intra-urban research. It views cities as complex systems and suggests that properties of cities can evolve non-linearly with city size[25–27]. Some studies have discussed in particular the power-law relationship between city population $N$ and area $A$ (the two determinants of density) $A \sim N^\gamma$, which is called "fundamental allometry" by some researchers[28]. Although there is a consensus that $\gamma < 1$ (large cities have a higher population density), empirical investigations show very diverging results, with $\gamma$ generally ranging between 2/3 and 1[10,29–33]. These discrepancies might be due to a major limitation of this literature, which does not consider the heterogeneities within cities. Indeed, the question of city definition is crucial here, because peripheral areas (where population density is low and which can be included or not depending on the definition) have an important effect on the overall density[34].

The primary objective of this work is to bridge intra-urban and inter-urban research, by investigating the evolution of the population density center-periphery structure ("radial profile") across population sizes. There are some precedents at a European scale[31,35], but the density radial profiles are not systematically modeled. We provide here the first comprehensive scaling analysis of the intra-urban structure of density, on a global sample of cities. We observe that population density decreases exponentially from the center. Despite a great variety of city sizes, development, location on the planet, we observe striking regularities in the radial structure across city sizes. The main fluctuations around the expected behavior can be explained by the level of development.

## Results

### Rescaled population density radial structure

Worldwide population is very unevenly distributed on the planet, as most people concentrate in cities. However, cities have very different population sizes, which makes it challenging to compare their urban structure[36]. This can be seen with maps. At a similar zoom level, it is very difficult to find consistencies between cities of different sizes (Fig 1a, thin edge). The main reason is that large cities like Tokyo ($N \simeq 37.4$ million inhabitants) are more spread out and tend to be denser than smaller cities like Sarajevo ($N \simeq 0.3$ million inhabitants), as suggested by the above literature.

We denote $\rho(r)$ the average population density as a function of distance to the center $r$. In order to remove the city size effect, we rescale the maps with the power one third of city population, $N^{1/3}$. In detail, we dilate the Euclidean space through a zoom on the map $r' = rk$, simultaneously with the vertical dimension of density $\rho' = \rho k$. The dilation factor $k$ which allows us to rescale any city, of population $N$, to the size of Tokyo (the largest city of our sample, taken as a reference without loss of generality), can be expressed as $k = (N_{Tokyo}/N)^{1/3}$. This approach takes off the size effect, leading to more homogeneous population density patterns across cities (Fig. 1a, thick edge). This implies that as cities grow, they experience two simultaneous phenomena: a densification and an expansion toward the periphery. Even for a city with shape distortions like Wellington, which is surrounded by hills and its harbor, the homothetic transformation remains very effective.

To obtain a more comprehensive view of the radial ("center-periphery") structure of cities, we turn the maps into population density radial profiles (Fig. 2a). Whatever the size, the population density profile is well-described by a negative exponential function, as straights lines appear on a semi-logarithmic graph. We sometimes observe a "density crater", which is already well-documented in the literature [8,14], as well as some cities with dense peripheral areas (nearby cities), such as Mexico or Beijing. Despite these peculiarities, curves are well clustered by city size at every distance from the center, indicating that large cities tend to be denser and cover a larger area to accommodate the extra population.

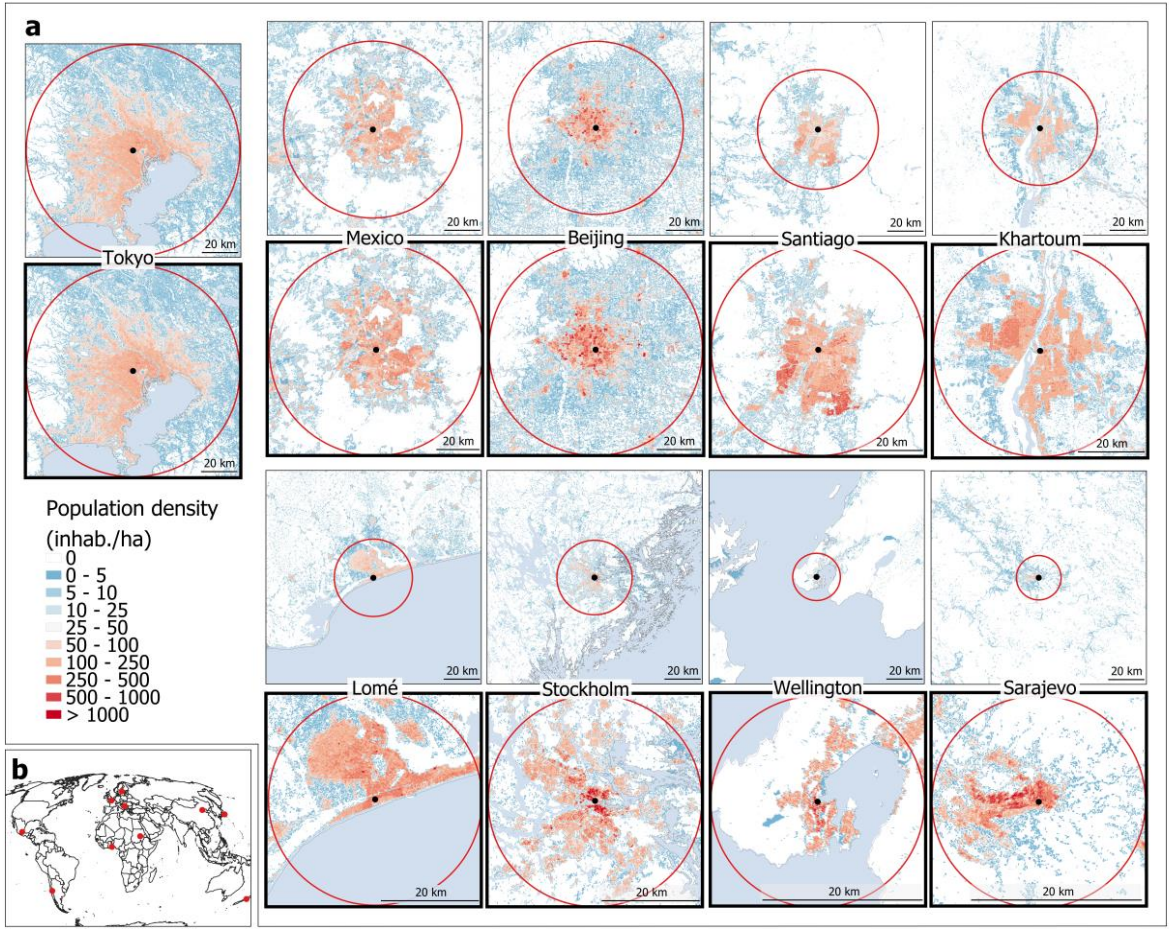

**Fig. 1 | View of population density in capital cities of different sizes. a** Population density in Tokyo (reference city, $N = 3.7 \times 10^7$ and $k = 1$), Mexico and Beijing ($N = 2.2 \times 10^7$ and $2.0 \times 10^7$; $k = 1.1$ and $1.2$), Santiago and Khartoum ($N = 6.8 \times 10^6$ and $5.8 \times 10^6$; $k = 1.8$ and $1.9$), Lomé and Stockholm ($N = 1.8 \times 10^6$ and $1.6 \times 10^6$; $k = 2.7$ and $2.8$), Wellington and Sarajevo ($N = 4.1 \times 10^5$ and $3.4 \times 10^5$; $k = 4.5$ and $4.8$). The circles represent a radius $r' = 60$ km. Top maps of each city are at the same scale (square of side 120km), while a rescaling has been performed on the bottom maps (thick edge), both on distances $r' = r.k$ and density $\rho' = \rho.k$. The city center is located by the black dot. Scale bars indicate $r = 20$km. **b** Location of these capital cities on the planet.

Following the same approach we used for the maps, we rescale the two Euclidean dimensions of space and the vertical dimension of population density. It allows us to get a view of the intra-urban density structure of cities, for a typical city of the size of Tokyo (Fig. 2b). We notice that our sample of capital cities exhibits similar variations of rescaled density $\rho'(r')$ despite their different history, size and location on the planet. These regularities indicate that the rescaling with the parameter $N^{1/3}$ is successful in taking off the size effect and point to a universal structure of cities arising from similar mechanisms. These findings are supported by robustness checks, which suggest that this 1/3 scaling exponent yields the most homogeneous rescaled radial profiles (Supplementary Information).

To further document the generic radial structure of all cities and the deviations, we compute the median rescaled radial profile and the interquartile range (IQR) with a spacing of 1 km around the center (Fig. 2c), which show a remarkable concentration of radial profiles around a mean global urban profile. To account for the unequal distribution of city sizes[36,37], we also compute those statistics using two additional weighting schemes, the Person Model (PM) and the P² model (P²M), which weight cities with city population and the square of city population, respectively (see Methods). Setting aside the central crater, the density at the center is $\rho'(r' = 0) \simeq 40{,}000$ inhab./km² for a theoretical city of the size of Tokyo (Fig. 2c), and drops sharply within the first kilometers from the center. By contrast, our scaling law of population density predicts a density at the center $\rho'(0) \simeq 27{,}000$ inhab./km² for a city of 10 million inhabitants, and $\rho'(0) \simeq 12{,}500$ inhab./km² for a city of one million inhabitants. These figures represent one of the first estimates of central density in relation to city population, since this size effect was first suggested[10]. We also note that both the median profile and the IQR remain nearly unchanged under the different weighting schemes, denoting that our approach is very robust along the urban hierarchy. This is confirmed by a look at the mean rescaled profiles of different city size classes, showing very small remaining fluctuations (Supplementary Information).

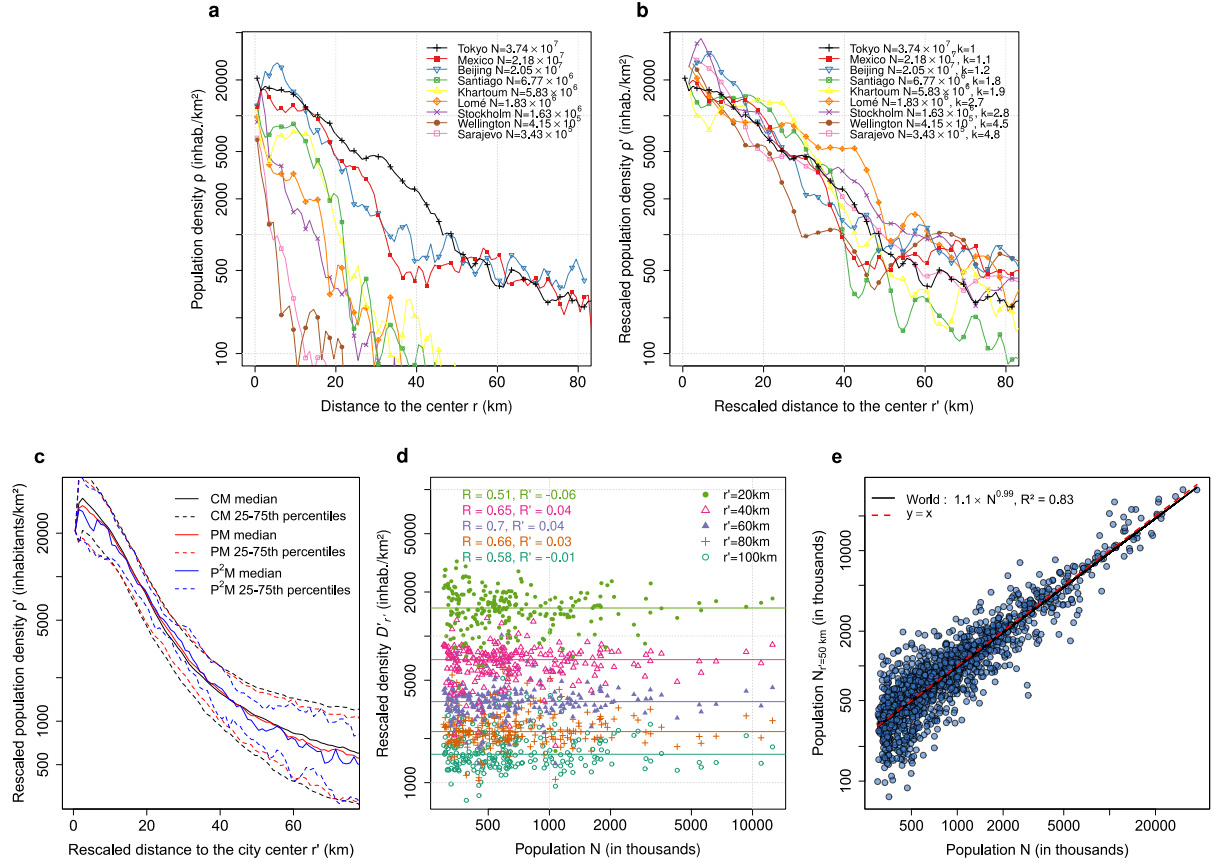

**Fig. 2 | Population density radial structure.** Radial profiles for a sample of nine capital cities displayed both before (**a**) and after (**b**) rescaling, on a semi-logarithmic plot. The population $N$ provided by the UN is given in the legend, as well as the rescaling factor $k$ (s.t. $r' = rk$ and $\rho' = \rho k$). **c** Distribution of population density radial profiles $\rho'$ as a function of distance to the center $r'$ for all cities of the database and under different weighting schemes. **d** Average rescaled density $D'$ in discs of radius $r'$ as a function of city size $N$. To ensure readability, results are displayed for European cities only (the Supplementary Information provides the same analysis for all cities). The correlation coefficient between city size and population density (before and after rescaling, written respectively R and R') is also given. Note that both variables are presented on logarithmic scales. The horizontal lines represent the median rescaled density $D'_{r'}$ of cities. **e** Population within a disc of radius $r' = 50$km as a function of UN population. The solid line is a power-law fit (equation (1)) with an exponent $0.99 \pm 0.01$.

These regularities in the radial structure of cities also suggest to look at them through circular discs. For simplicity, we will speak here of rescaled distance $r'$ to the center, actually meaning a non-rescaled radius $r = r'/k$ which differs for each city (see the circles on Fig. 1a). Given a value of the rescaled radius $r'$, we define $N_{r'}$ as the population count, $A_{r'} = \pi r'^2/k^2$ the area, and $D_{r'} = N_{r'}/A_{r'}$ the density within a disc of rescaled radius $r'$ around the city center. The rescaled density is given by $D'_{r'} = D_{r'} \times k$, where $k$ is the rescaling factor defined above. We test different values of $r'$, in order to look at the scaling effect from the urban core ($r' = 20$km) to the whole city ($r' = 100$km). Urban density decreases as (rescaled) distance $r'$ from the center increases, because more peripheral areas are encompassed (Fig. 2d). While we find a clear positive correlation between density $D_{r'}$ and city population $N$ ($p$-value <$10^{-16}$), this correlation vanishes when using the rescaled density $D'_{r'}$ (see Supplementary Information), indicating that the transformation holds throughout the urban system and along the entire radial profile. This is an important finding for the understanding of urban structures, as it reveals a rather universal shape of cities, which can only be seen if we account for the size effect.

Instead of examining density, we now focus on the fundamental relationship between area $A_N$ and population $N$, which is a common question in the urban scaling literature[34,38]. This fundamental allometry[28] is expressed by the power-law relation

$$A_N \sim A_1 N^\gamma, \qquad (1)$$

where $A_1$ is a constant corresponding to the theoretical area of a city with one inhabitant, and $\gamma$ a scaling exponent. However, no city delineation is provided by our UN WUP dataset[39]. Nevertheless, we draw concentric circles of rescaled radius $r'$ around the city centers, and look at the relationship between the population $N_{r'}$ and city size $N$. The scaling relationship for $r' = 50$ km returns an exponent $0.99 \pm 0.01$ (R² = 0.83), and the curve is very close to the straight line $x = y$ (Fig. 2e). This finding suggests that the definition used by the UN (using many criteria) is roughly equivalent, on average, to a circle of rescaled radius $r' = 50$ km, thus following the homothetic behavior of urban density. Then, if the largest city, Tokyo, can be delineated with a circle buffer of radius $r = r' = 50$ km, this distance $r' = 50$ km actually decreases with population, corresponding to $r \simeq 34$ km and $r \simeq 10$ km for a ten million and a one million inhabitants city, respectively. Since $A_{r'} = \pi r'^2/k^2 \sim N^{2/3}$ as $k \sim N^{-1/3}$, these results regarding population density determine a relationship between area and population of the form $A \sim N^{2/3}$. They also confirm the roughly circular shape of cities, featured in classic models of urban geography[40,41].

## Scaling of the parameters

We now aim to describe these radial profiles of population density mathematically with simple statistical tools, and a parsimonious approach. To achieve this, we use a limited number of parameters that can be directly interpreted in terms of urban concentration. As suggested by the previous results, we carry out exponential fits for each city. We use a linear fit L of the logarithm (equation (2), see Methods), as well as a non-linear fit of the raw value (NL model, equation (3)), with two parameters, $a_N$ measuring the density at the center and $b_N$ the characteristic distance. Besides, we try another non-linear (exponential) fit, where we force the relationship $a_N b_N^2 2\pi = N$ to hold (see Methods), ensuring that the total population in the model corresponds to our UN data. Using the relationship between the two parameters $a_N$ and $b_N$ (equation (5), see Methods), this fit is a one-parameter model (1NL model). For comparison purposes, we also fit the profiles using a (negative) power law.

We notice that the exponential function describes empirical data satisfactory in most world cities, with median R² values of 0.52, 0.95 and 0.91 for the linear, non-linear and one-parameter non-linear fits, respectively (density plot, Supplementary Information). These values are high, which suggests that very parsimonious models are able to describe population density profiles. By contrast, the power-law function is clearly outperformed by the exponential (Supplementary Information).

Let us call $\alpha$ and $\beta$ the scaling exponents of the fitting parameters $a_N$ and $b_N$ (equation (7)). From the 3-dimensional homothetic scaling with the exponent 1/3 observed before, we expect $\alpha = \beta = 1/3$. This is rather precisely what we find on Fig. 3, where the expected exponent 1/3 is represented by a horizontal dashed line. For nonlinear models, we note that a weighting scheme which distributes statistical weights more evenly along the urban hierarchy (see Methods), such as the Person and P² models (PM and P²M), brings us closer to these expectations. Conversely, if small cities are weighted as much as megacities (City model CM), we find a higher value for $\beta$ (horizontal scaling) than for $\alpha$ (vertical scaling).

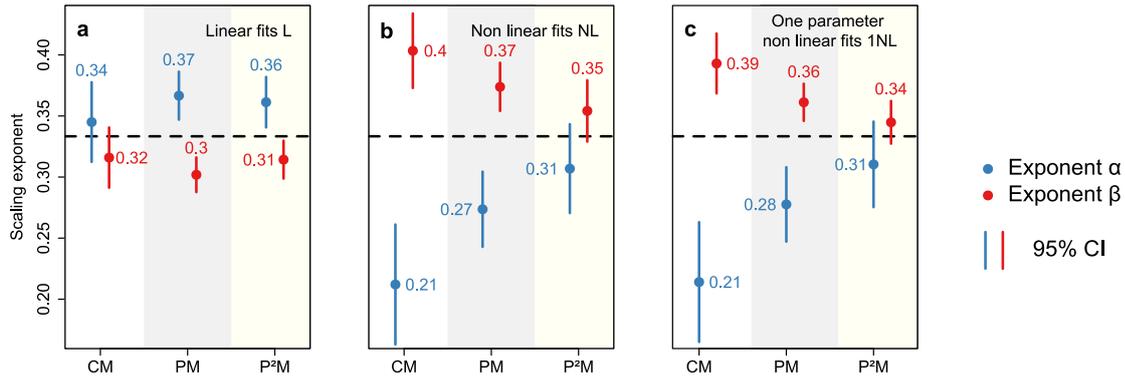

**Fig. 3 | Scaling analysis of the fitted parameters.** Results of linear fits of the fitted (log) parameters $a_N$ (exponent $\alpha$) and $b_N$ (exponent $\beta$) against the (log) total population $N$. Those parameters $a_N$ and $b_N$ come themselves from the linear fits (**a**), as well as two-parameters (**b**) and one-parameter (**c**) non-linear fits of the radial profiles. We derive these results for three weighting possibilities: the City Model (CM), the Person Model (PM) and the P²M model. The horizontal dashed line represents the expected value $\alpha = \beta = 1/3$. The error bars indicate the 95% confidence interval (CI).

We also observe that this scaling behavior is consistent with another theoretical expectation. If cities are circular and population density decreases exponentially with the distance to the center, then the total population can be seen as the volume under the radial profile of density, and a scaling relationship $a_N \, b_N^2 \sim N$ holds (see Methods). Hence, the two scaling parameters are connected by the equality $\alpha + 2\beta = 1$ (equation (5)), which aligns with the exponents obtained from the power-law relationships (Fig. 3), in particular the homothetic scaling $\alpha = \beta = 1/3$.

In light of our findings regarding population density, cities of the world are roughly of one shape, exponential cones which evolve vertically and horizontally with city size. Therefore, the characteristic distance and central density for a city of any size can be expressed as $b_N = b_1 N^{1/3}$ and $a_N = a_1 N^{1/3}$, respectively, where we estimate $b_1 \simeq 35$ m and $a_1 \simeq 125$ inhab./km² using the results of the 1NL model. Fig. 4a provides a theoretical display of the average change in the radial structure of population density, as we move from a small to a large city. Of course, individual cities do not follow precisely these geometric properties, but this lays the foundations for more effective inter-urban comparisons.

Indeed, cities have local characteristics likely to infringe on the expected radial structure. The scatter plot presented in Fig. 4b illustrates the relationship between the fitted parameter $b_N$ (from the 1NL model, Fig. 3c) and city size, along with the fluctuations around the scaling law. Although many cities exhibit a center-periphery structure close to the expected homothetic transformation, a closer inspection reveals significant deviations. Some capital cities are rather sprawled, as most of their population is located in peripheral areas (Brasilia, Canberra, Washington D.C), while some other cities are much denser (Dhaka, Cairo). We observe that these different urban forms are linked to the nationwide energy use per capita. Compact cities (with a low value of $b_N$ considering their population) are located in energy-efficient countries, while sprawled cities show the opposite trend. To further investigate regional and national peculiarities, one option is to account for the city-size effect beforehand. Our approach differs in this respect from many studies, which recognize that urban forms vary with city population, but do not account for this effect systematically.

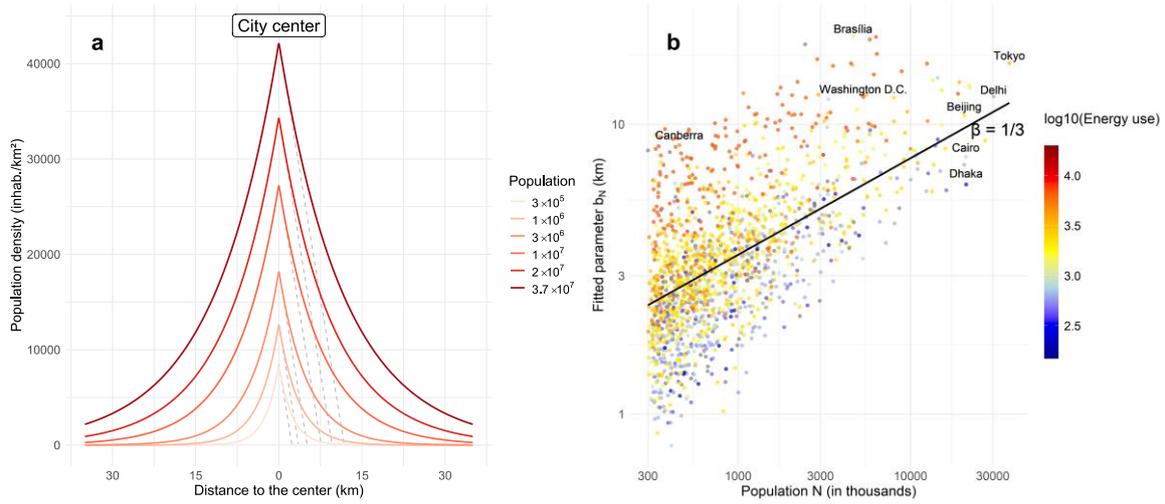

**Fig. 4 | Urban scaling of population density radial profiles. a** Illustration of the homothetic behavior of population density radial profiles. The tangent at $x = 0$ (grey dashed lines) is also shown. **b** Power-law relationship connecting $b_N$ (from the one-parameter non-linear fit 1NL) and city size $N$. The black line gives our expectations, a scaling with the cube root of total population $N^{1/3}$. The color represents the annual energy use per capita, in kilograms of oil equivalent (in decimal logarithmic scale).

## Urban density index

Although our urban scaling framework manages for the city size effect surprisingly well, we still observe significant fluctuations across different locations on the planet (Supplementary Information). Our results suggest a size-independent measure of urban concentration, the urban density index $\text{UDI} = b_N/N^{1/3} = (b_N/a_N)^{1/3}$ (see Methods). It corresponds to the residual of the scaling relationship between the parameter $b_N$ and population size $N$ (Fig. 4b) and is linked to average density $\rho$ as $\rho \sim UDI^{-2}$ (see Methods), so that a city gets denser as the UDI decreases (Fig. 5a). Each population size $N$ associates indeed with a range of values for the two parameters $a_N$ and $b_N$, and for the UDI. This index provides a more in-depth comparison between cities by examining how the center-periphery structure of cities deviates from the observed scaling law. It has the dimension of a distance, but could also be considered dimensionless (see Discussion).

The map and boxplot illustrate the geographical variation in urban concentration measured by the UDI (Figs. 5b and 5c). It reveals a specific distribution, statistically close to log-normal (Fig. 5d), which mainly seems to oppose developed to developing countries. The latter countries (such as Bangladesh, Egypt, India) gather very dense cities, while sprawled cities are mainly located in high-income countries (United States, Australia, Canada). However, there are some exceptions. We note that South African cities have a very flat center-periphery structure, which can be explained by the stringent urban regulations during Apartheid[13]. We further observe that there are rather small variations within countries, suggesting similar mechanisms at national level. Among the global driving forces, wealth measured by income per capita seems to be key, as low-income and lower-middle income countries concentrate compact cities (low value of the UDI), and conversely for high-income countries (Fig. 5b).

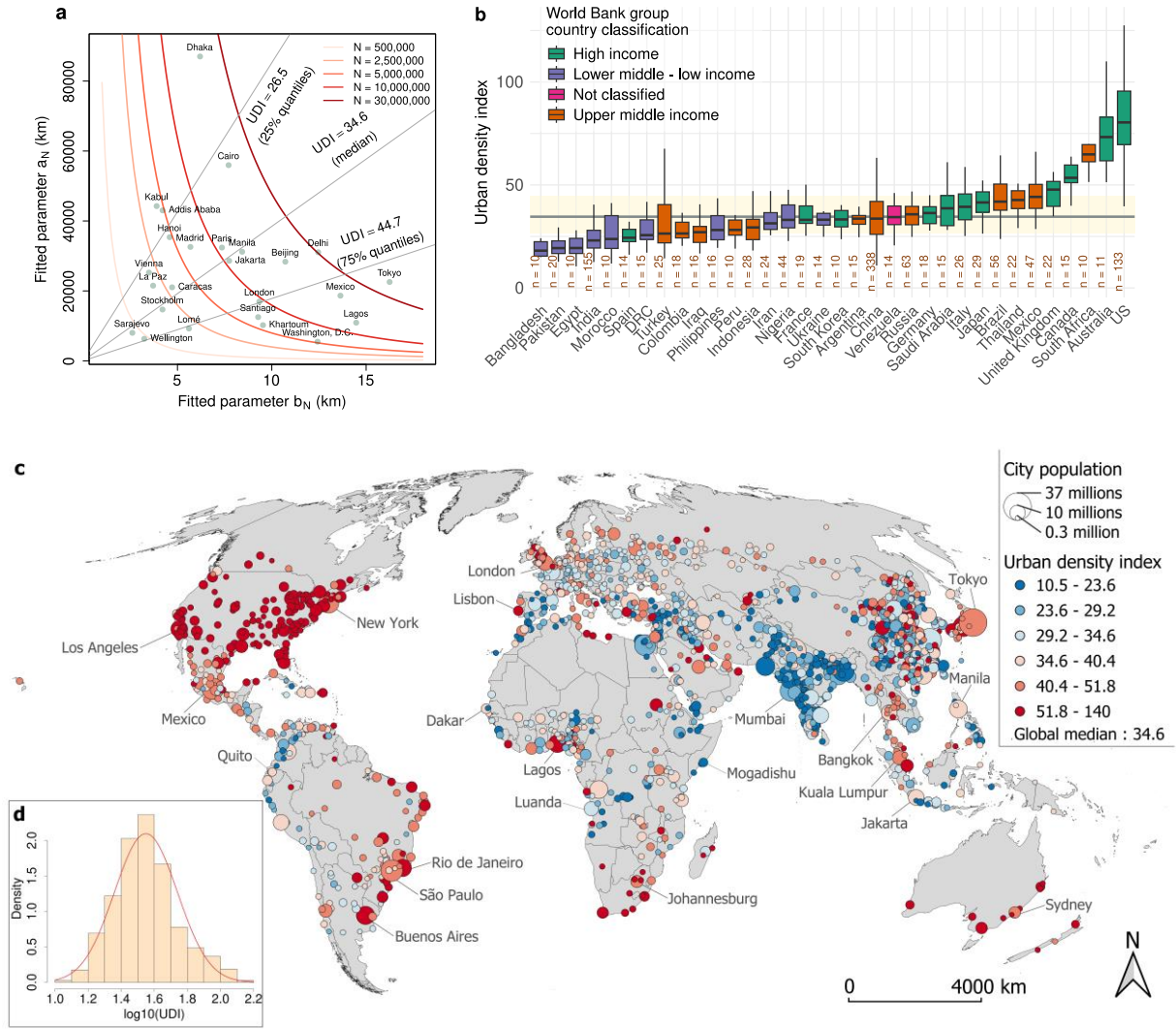

**Fig. 5 | Urban density index UDI. a** Relationship between the regression parameters $a_N$ and $b_N$ of the 1NL model and the UDI, for different population sizes. The dots represent the estimated parameters for a selection of capital cities. **b** Boxplot of the UDI**.** Results are presented by country and sorted by median value (increasing order, from left to right). The number of cities per country $n$ is also given (countries having fewer than 10 cities are not displayed). The color indicates the income group (World Bank). The horizontal line indicates the global median, $UDI = 34.6$, and the colored area the 25%-75% quantiles. **c** Map of the UDI. **d** Histogram of the logarithm of the UDI. A normal distribution of the same mean 1.5 and standard deviation $\sigma = 0.19$ (red curve) is used as a guide to the eye.

We deepen this analysis by exploring how the UDI evolves with a set of indicators obtained at national level (Supplementary Information). Our results show that two variables are even more closely linked to the UDI than the GDP per capita, namely energy use and number of cars per inhabitant – these are all indicators of the level of development. The interplay between UDI, population density and energy use per capita can be related to previous work[42], which finds a negative relationship between average population density $\rho$ and energy use per capita $\varepsilon$, and pleads for more compact cities in order to meet the sustainability challenge. Indeed, we have $UDI \sim \varepsilon^{0.4}$ and $\rho \sim UDI^{-2}$, which gives us $\varepsilon \sim \rho^{-1.3}$ (Supplementary information).

## Discussion

Describing urban population density $\rho_N(r)$ as a function of the distance $r$ to the city center and total population $N$, $\rho_N(r) \sim \frac{N^{1/3}}{2\pi b_1^2} \cdot \exp\left(-\frac{r}{b_1 N^{1/3}}\right)$, with only one parameter $b_1 \simeq 35$ m, we uncover here what can be considered as the first global result regarding the spatial internal structure of cities. In that sense, we could relate it to Zipf's law, which has been widely studied and observed, but does not consider the internal structure of cities. This result both gives another benchmark on which to build urban science (for instance, to tackle the still open problem of city definition), and raises many questions and research perspectives. First, as for Zipf's law, to what extent does it hold at the lower end of the size distribution, where effects of polycentricity, and issues of city definition, are more critical?

In economic terms, we find no economies or diseconomies of scale when it comes to population density; instead, we find constant returns to scale or, in other terms, an isometric scaling or homothety. Why do cities present such a simple geometrical scaling? It is true that there is huge cultural, climatic, political, historical, landscape, social heterogeneity in the world. But still, there should be more common ground than just the 3-dimensional geometry of Euclidean space. And why is the exponential shape so robust? The pioneering work of Nordbeck[29] discusses this universal shape of cities (a "volcano" in his words) and suggests that a homothetic scaling would appear visually appealing and geometrically logical, but this is not a satisfactory explanation. We see here clearly that population density is a variable which behaves like a volume. The two horizontal dimensions $x$ and $y$ are geographic and mixed in the radial distance $r = \sqrt{x^2 + y^2}$; therefore, it is rather trivial that they have similar scaling exponents (and hence the $a_N b_N^2 2\pi = N$ relationship of equation (5), which identifies total population $N$ to a "volume" of population). On the other hand, the vertical dimension of population density stands out since it is not really physical, as it involves human beings. This puzzle can be addressed with a simple hypothesis that all inhabitants of a city occupy a roughly similar building volume, which converts the vertical aspect of population density into a physical measure of vertical distance (and makes the UDI dimensionless). While it may seem unusual to combine human residential choice behavior with such purely geometric considerations, our findings bring us to this point.

In order to test this hypothesis, a research perspective is the study of buildings and their volumes, with similar methods as those used here. Additionally, in order to know more precisely how physical the vertical dimension of population density is, it would be interesting to have access to 3-dimensional data on people's locations at various times to understand how high above (or below) ground level city residents live, alongside their horizontal geographic location. No such data is available yet to our knowledge, but it might appear in the future (likewise, the accuracy of global population density layers will presumably increase, making such research more precise). It can be anticipated that the average 3-dimensional distribution of urban populations will follow a homothetic volume-scaling law similar to the one identified here. Additionally, comparable studies focusing on employment density (rather than residential population density) and commuting patterns would also be interesting steps for improving our understanding of cities, their structure and dynamics.

Naturally, these empirical results also raise questions on the urban modeling side. By linking intra-urban analysis (inspired in this by theoretical models from urban economics) and inter-urban analysis through the urban scaling framework, we enlighten the still open problem of the most basic allometry of city population $N$ versus area $A$, $A \sim N^{\gamma}$ [34]. With a scaling of population density profiles in the cube root of total population $N^{1/3}$ in the two horizontal geographical dimensions of space, we find that this fundamental allometry has an exponent $\gamma = 2/3$ at the global scale, $A \sim N^{2/3}$. This result, rejoining much earlier ones[29], is a fundamental input in urban modeling approaches[28,38], but as such, there is no attempt to explain it. And urban economic approaches, which offer explanations for the internal spatial structure of cities, lack a realistic description of space, particularly in terms of land use, revealing a modeling gap here[43,44].

Regarding the sustainability of our urban world, these results highlight the positive relationship between urban extent —measured by our size-independent urban density index UDI— and per capita wealth, income, and especially energy use and car ownership. This is also coherent with the link at the micro-level between wealth and environmental footprint[45]. We rejoin here, with different methods and data, much discussed earlier work[42], which observes a negative relationship between average population density $\rho$ and energy use $\varepsilon$. This is also consistent with the link between income and sprawl in the standard urban economic model[13,23]. In a world where fossil fuel reserves, formed over millions of years, are rapidly depleting within a few decades, along with the metals required to replace current internal combustion engine vehicles with electric ones, urban planning towards global sustainability should clearly aim at low UDI. This involves promoting dense, compact cities with reduced car dependency and lower energy consumption per capita, similar to those in lower-income countries. Of course, this could also lead to negative effects, such as higher housing prices. The link between our results on the distribution of urban population and housing prices could actually be an interesting perspective of research, introducing more human behavior, through the housing market, to understand the very geometric scaling observed here.

Our work opens up new avenues for future investigations, since it provides a powerful tool to compare cities with each other. Future research could explore national and continental differences more systematically, in order to gain deeper insights into the factors shaping urban forms. More broadly, we highlight the importance of the intra-urban structure of cities in understanding patterns of human settlement and designing policies that promote sustainable development. Our study contributes to document the multi-dimensional impacts of density and suggests that we should aim for compact urban development—however, we must remember that true change mostly relies on a collective will.

# Methods

### Data collection

Various methodologies have been proposed to provide spatially disaggregated population estimates, but this task remains challenging, especially if census data is lacking or inaccurate[46]. Here, we use the latest version of the GHS-POP layer, which provides residential population estimates in 2020 across the globe, at a resolution of $100 \times 100$m[47]. It is produced through a dasymetric top-down approach, where census population from the Gridded Population of the World, version 4.11 (GPWv4.11) is disaggregated to grid cells, proportionally to built-up volume[48]. We note that this approach is rather simple and consistent globally. On the other hand, however, estimates at pixel level are subject to some uncertainties due to various factors, including the quality of census data and the stable relationship between population and built-up volume[49]. Recently, some authors have proposed a review of large-scale population grids, discussing their methodological approach, resolution, spatial coverage and fitness for use[50].

We use additional datasets to observe the relationship between socio-economic variables and the population density radial structure. In detail, we gather a set of indicators at country level from various institutions. In the Supplementary Information, we provide a detailed description of these datasets (source, description, reference year).

### Sample of cities

Our sample includes 1,860 cities in the word with 300,000 inhabitants or more in 2018 from the World Urbanization Prospects (WUP), and gathering some 2.7 billion inhabitants in 2020[39]. Population estimates for each city at 5-year interval between 1950 and 2035 are also provided. For our reference year 2020, Tokyo was the most populated city (37.393 million inhabitants).

From this initial sample, we remove some cities which can be considered as subcenters of a larger adjacent urbanized area. For full details, one may refer to the Supplementary Information. We end up with 1593 cities worldwide. The most represented countries are China (338 cities), India (155), the US (133), Russia (63) and Brazil (56).

### Population density radial structure

We compute population density as a function of the distance to the city center, whose location is chosen to be the city hall. If not available, we choose instead a central building, such as the central post office or the key train station in India, or the central mosque in the Middle East and North Africa. Thereafter, we reproject the population grid in the local UTM projection and draw concentric rings with a spacing of 1 km around the city center. Each GHS-POP cell is allocated to a ring, taking as a reference the center of the cell. We finally compute the mean population density within each ring. Because large cities have a further-reaching influence than small cities, the population density is computed up to the distance $r'_f = r_f \times (N_{Tokyo}/N)^{1/3} = 100$ km, where $r_f$ and $N$ are the maximal radius and the population of a given city, and $N_{Tokyo}$ the population of Tokyo. Hence, the maximal distance $r_f$ scales with population as $r_f \sim N^{1/3}$.

Thereafter, we carry out exponential fits. We perform both a linear fit of the logarithm (L) and a non-linear fit of the raw value (NL), respectively following

$$\log(\rho_N(r)) \sim \log(a_N) - \frac{r}{b_N}, \tag{2}$$

and

$$\rho_N(r) \sim a_N \cdot \exp\left(-\frac{r}{b_N}\right), \tag{3}$$

where $a_N$ is the (theoretical) population density at the center, $r$ the distance to the center (in kilometers), and $b_N$ the distance at which population density has been divided by $e^1 \simeq 2.7$ (it is also the distance at which the highest share of total population can be found, and half of the mean distance to the center of all individuals, see Supplementary Information). Note that both L (equation (2)) and NL (equation (3)) regressions yield different values of the parameters $a_N$ and $b_N$. Indeed, while the linear fit minimizes the relative error, the non-linear fit minimizes the absolute error, which has an impact on the estimated parameters[51]. Further, we perform a one-parameter non-linear fit (1NL), making use of the fact that integrating the radial profile of density should yield the total population $N$. Supposing an exponential radial profile of population density, the cumulative population $N_r$ in a disc of radius $r$ around the city center can be written as

$$N_r = \int_0^r a_N \cdot exp\left(-\frac{x}{b_N}\right) \cdot 2\pi x \, dx = 2\pi a_N b_N^2 \left[1 - \left(1 + \frac{r}{b_N}\right) \cdot \exp\left(-\frac{r}{b_N}\right)\right] \tag{4}$$

When $r$ is large compared to the characteristic distance $b_N$, we obtain

$$\lim_{r \to +\infty} N_r = 2\pi a_N b_N^2 = N, \tag{5}$$

which has already been stressed some decades ago in the early studies of urban population densities [8,22]. We note that the maximal extent $r_f$ respects this condition well since $r_f/b_N \simeq 9$ in our sample on average, meaning that the theoretical value $N_{r_f}$ given by equation (4) is only 0.1% smaller than $N$. By using equation (5), we can rewrite equation (3) as

$$\rho_N(r) \sim \frac{N \cdot \exp\left(-\frac{r}{b_N}\right)}{2\pi b_N^2}, \tag{6}$$

where $b_N$ is the only parameter to be estimated. Central density $a_N$ is directly given by $a_N = N/b_N^2 2\pi$, thanks to equation (5).

## Scaling of urban profiles

We expect the population density at the center $a_N$ and the characteristic distance $b_N$ to scale with city size $N$. Once radial profiles are fitted, we relate the estimated parameters $a_N$, $b_N$ (from both linear and non-linear fits) to population size $N$ via power-law functions

$$a_N \sim a_1 N^\alpha \text{ and } b_N \sim b_1 N^\beta, \tag{7}$$

where $\alpha$ and $\beta$ are two scaling exponents, $a_1$ the theoretical density at the center and $b_1$ the characteristic distance for a city of $N = 1$ inhabitant. We estimate the scaling parameters $\alpha$ and $\beta$, with linear fits of the logarithms of $a_N$, $b_N$ and $N$.

To check if the scaling relationship holds throughout the system of cities, we derive the analysis with different weighting possibilities. The usual City Model (CM) weights all cities equally. In addition, as prescribed by earlier studies[37], we try the Person Model (PM), which weights cities proportionally to their population $N$. We also add a third possibility, namely the P² Model (P²M), which weights cities with the square of city population, $N^2$. This last weighting scheme is able to cancel the distribution of cities' sizes, as the cumulative probability density associated with Zipf's law[36] $P(x > N) \sim 1/x$ gives a probability density function $P(x = N) \sim 1/x^2$, thus ensuring an homogeneous distribution of weights at all scales using the P²M model.

## Urban density index

The scaling of urban profiles suggests to use a size-independent measure of urban concentration, defined using the parameters of the 1NL model. We introduce the urban density index $UDI = b_N/N^{1/3}$, which corresponds to the residual of the regression displayed on Fig. 4b (see Supplementary Information). Hence, two cities $i$ and $j$ with different population sizes are equal in UDI if $b_i/N_i^{1/3} = b_j/N_j^{1/3}$ holds, or equivalently $b_i/a_i = b_j/a_j$ (Fig. 5a). This index measures how spread a city is, since $b_N \sim UDI$ and how dense it is in the center, since $a_N \sim UDI^{-2}$ (using equation (5)), both independently of its size $N$. Thus, every 1% increase in the index associates with a 1% increase in the characteristic distance $b_N$ and a 2% decrease in density at the center $a_N$. Since equation (5) holds, we also note that the UDI can be computed as $(b_N/a_N)^{1/3}$, which is linked to the inverse slope of the tangent in Fig. 4a, whose equation is

$$y = -\frac{a_N}{b_N}x + a_N. \tag{8}$$

The index synthetically summarizes whether a city is compact and dense or spread out, regardless of its size (Supplementary Information). If a city has a low UDI value, its population is more densely concentrated near the center. Conversely, a high value of the UDI indicates that city population is spread out.

**Data availability**

The population grid at 100 m resolution that is used in this study is available in the European Commission's Joint Research Centre Data Catalog, and can be accessed at https://human-settlement.emergency.copernicus.eu/. Population of largest urban agglomerations (300,000 inhabitants or more) can be downloaded on the United Nations' website, at https://population.un.org/wup/downloads. National socio-economic indicators used in this study are described in the Supplementary Material.

**Code availability**

The geospatial data analysis was programmed in R 4.2.2 and using standard packages. The code used for analyzing data is available from the corresponding authors on request.